\documentstyle[aps,prc,twocolumn,floats,epsfig]{revtex}

\tighten

\newcommand{\half}[1]{{\scriptstyle{\frac{#1}{2}}}}
\newcommand{\nuc}[2]{{{$^{#1}$#2}}}

\begin{document}
\twocolumn[\columnwidth\textwidth\csname@twocolumnfalse\endcsname

\title{Fine Structure in the Decay of Deformed Proton Emitters:
Non-adiabatic Approach}
\draft
\author{A.T. Kruppa,$^{1,2}$
  B. Barmore,$^{2,3}$
  W. Nazarewicz,$^{3-5}$ and
  T. Vertse$^{1,2}$ }

\address{$^{1}$Institute of Nuclear Research of the Hungarian
  Academy of Sciences, P.O. Box 51, H-4001, Debrecen, Hungary}
\address{$^{2}$Joint Institute for Heavy Ion Research, Oak Ridge
  National Laboratory, P.O. Box 2008, Oak Ridge, Tennessee 37831}
\address{$^{3}$Department of Physics and Astronomy, University of Tennessee,
  Knoxville, Tennessee 37996}
\address{$^{4}$Physics Division, Oak Ridge National Laboratory, P. O.
  Box 2008, Oak Ridge, Tennessee 37831}
\address{$^{5}$Institute of Theoretical Physics, Warsaw University,
  ul. Ho\.{z}a 69, PL-00681, Warsaw, Poland}

\maketitle

\begin{abstract}
The coupled-channel Schr\"odinger equation with outgoing
wave boundary conditions is
employed to study
the fine structure  seen in the proton decay of deformed
even-$N$, odd-$Z$ rare earth nuclei $^{131}$Eu and $^{141}$Ho.
 Experimental lifetimes and proton-decay branching ratios
are reproduced. The comparison with the standard adiabatic theory
 is made.
\end{abstract}


\addvspace{5mm}]  

\narrowtext

Proton radioactivity  has proven to be a very powerful
tool to observe
neutron-deficient nuclei and study their structure.
Theoretically, proton radioactivity is an
excellent example of a simple  three-dimensional
quantum-mechanical tunneling problem. Indeed, in  first order,
it only involves a single
proton moving through  the Coulomb
 barrier of the daughter nucleus. In reality, the process
of proton emission
is more complicated since the perfect separation of the
nuclear many-body wave function into that of the proton and
the daughter cannot be made, and -- in addition --
  the decay
is greatly influenced by  nuclear structure effects such
as configuration mixing.
In spite of this, the first-order
one-body picture works surprisingly
well; it enables us to determine the angular momentum
content of a resonance and the associated spectroscopic factor
in many cases
\cite{[Hof95b],[Woo97]}.
Experimental and theoretical
investigations of proton emitters
are  opening up a  wealth of
exciting physics associated with the
 coupling between bound states and extremely
narrow resonances  in  the region of very low
single-particle level density. One particular example of such a
coupling,
due to the Coriolis interaction, is discussed in this work.

The last two years have seen an explosive number of
exciting  discoveries in this field, including new ground-state
proton emitters and  proton-decaying excited states
\cite{[Dav98],[Bat98],[Ryk99],[Bin99]},
and the first evidence  of fine structure in proton
decay \cite{[Son99]}.
The main focus of recent
investigations has been on well-deformed systems which
exhibit collective rotational motion; consequently, they are splendid
laboratories for the interplay between proton
emission and angular momentum.

{} From a theoretical viewpoint, the understanding of proton
emitters is a test of how well   one can describe very
narrow resonances.
For spherical nuclei,
there are many available theoretical methods,
most of which give very
similar and accurate results~\cite{[Abe97]}.
There have been several theoretical attempts to describe
deformed proton  emitters.
These approaches can be divided into three groups.
The first family of calculations
\cite{[Dav98],[Son99],[Bug89]} is based on the
reaction-theoretical framework of Kadmenski\u{\i} and
collaborators \cite{[Kad73]}. The second group
is based on the theory of Gamow (resonant) states
\cite{[Ryk99],[Fer97],[Mag98],[Mag99]}. Finally,
calculations based on  the
time-dependent Schr\"odinger
equation have recently become available \cite{[Tal98]}.
 All of these papers
assume a strong coupling approximation. That is,  the daughter
nucleus is considered to be a perfect rotor with an infinitely large
moment of inertia. Consequently, all the members of the
ground-state rotational band are  degenerate
and the Coriolis coupling is ignored. Our work
is the first attempt to go beyond these simplified assumptions in
the description of proton radioactivity.

Our technique is based on the theory of Gamow states.
More precisely, we
solve the coupled-channel Schr\"{o}dinger equation
describing the motion of the proton in the
deformed average potential of a core (a daughter nucleus).
It is assumed that the wave function of the proton is regular at
the origin and asymptotes to a purely outgoing Coulomb wave.
These boundary conditions
result in complex-energy eigenstates \cite{[Hum61]}.
For resonant states, the real part of the energy,
$E_0={\rm Re}(E)$, can
be interpreted as the resonance's energy, while the imaginary
part is proportional to the resonance's width,
$\Gamma=-2{\rm Im}\,(E)$.

Let us consider the Hamiltonian of the daughter-plus-proton
system
\begin{equation}\label{Htot}
H=H_{d} + H_p + V,
\end{equation}
where $H_{\rm d}$ is the Hamiltonian of the daughter nucleus,
$H_p$ is the proton Hamiltonian, and $V$
represents the  proton-daughter interaction.
The total wave function, $\Psi$, of the parent nucleus
can be written in the weak-coupling form
\begin{equation}\label{Psi}
\Psi_{JM}=r^{-1}\sum_{J_dl_pj_p} u_{J_dl_pj_p}^J(r)
\left({\cal Y}_{l_pj_p}
\otimes \Phi_{J_d}\right)_{JM}.
\end{equation}
In (\ref{Psi})  $u_\alpha^J$
($\alpha$$\equiv$$(J_dl_pj_p)$ labels the channel quantum numbers)
is the cluster radial
function representing the relative radial motion of the proton
and the daughter nucleus, ${\cal Y}_{j_pl_pm_p}$ is
the orbital-spin wave function of the proton, and
$\Phi_{J_dM_d}$ is the wave function of the daughter
nucleus. By definition, one has
\begin{equation}\label{HPsi}
H_{d} \Phi_{J_dM_d} = E_{J_d} \Phi_{J_dM_d}.
\end{equation}
In practice, the energies $E_{J_d}$ are taken from experiment
or, if the data are not available, they are modeled
theoretically. Inserting (\ref{Psi})  into the Schr\"odinger
equation and integrating over all coordinates except the radial
variable $r$, one obtains the set of coupled equations for
the cluster functions \cite{[Bug89],[Tam65]}:
\begin{eqnarray}\label{cc}
  \left[-\frac{\hbar^2}{2\mu} \frac{d^2}{dr^2} \right.
& + & \left.\frac{\hbar^2 l_p(l_p+1)}{2\mu r^2} +
  V_{\alpha}(r) - Q_{J_d} \right] u_{\alpha}^J(r) \nonumber\\
 & + & \sum_{\alpha'} V_{\alpha,\alpha'}^J(r)
 u_{\alpha'}^J(r)=0.
\end{eqnarray}
In Eq.~(\ref{cc})
$V_{\alpha}$ represents the average spherical potential
of the proton in the state $\alpha$,
$V_{\alpha,\alpha'}^J$ is the off-diagonal coupling
term, and
 $Q_{J_d}$ is the energy of the
relative motion of the proton and daughter nucleus in the
state $J_d$. One  obviously has $Q_{J_d}=Q_0-E_{J_d}$,
where  $Q_0$ is the $Q_p$-value for the decay to the
$J_d^\pi$=0$^+$ ground state.

The method of coupled-channels  described above
has several advantages over the commonly used strong coupling formalism.
First,  excitations in the
core may be included in a straightforward manner.  This enables us to
study the  proton decay from the rotational bands of the parent nucleus
to  various rotational states of the daughter. Furthermore, since the formalism
is based on the laboratory-system description [Hamiltonian (\ref{Htot})
is rotationally invariant and the wave function $\Psi$ conserves angular
momentum], the Coriolis coupling is automatically included.

The coupled equations (\ref{cc}) are solved in the complex
energy plane. Asymptotically, the cluster wave function
$u_{J_dj_pl_p}^J$ behaves like a purely  outgoing Coulomb
wave $G_{l_p}(k_{J_d}r)+iF_{l_p}(k_{J_d}r)$ with
$k_{J_d}=\sqrt{2\mu Q_{J_d}}/\hbar$. In this work we assume
that the average single-particle potential is approximated
by the sum of a  Woods-Saxon (WS) potential, spin-orbit term,
and the Coulomb potential. The  axially deformed WS potential
is defined according to Ref.~\cite{[Cwi87]}. We employ
the Chepurnov parameterization  \cite{[Che67a]}; it
gives good agreement with proton single-particle energy levels as given in
Ref.~\cite{[Naz90]}.  The Chepurnov
parameterization  provides a reasonable compromise between
the  Becchetti-Greenlees parameter set \cite{[Bec69]}
(excellent for the description
of reaction aspects but slightly displacing the $h_{11/2}$, $g_{9/2}$,
and  $s_{1/2}$ proton shells) and the universal parameter set  \cite{[Dud81]}
(excellent for the description  of structure properties of deformed rare earth nuclei
\cite{[Naz90]} but  having too large a radius to give a quantitative
description of
the tunneling rate \cite{[Abe97]}).

Since the resonance  energy cannot be predicted with sufficient
accuracy, following Refs.~\cite{[Ryk99],[Abe97]},
the depth of the WS
potential is adjusted to give the experimental $Q_0$
value.   The deformed part of the spin-orbit interaction is
neglected; we do not expect this to have a noticeable effect on the
results~\cite{[Nil55]}.
 The off-diagonal coupling in  (\ref{cc})
appears thanks to the non-spherical parts of  WS and Coulomb
potentials.
The exact form of  $V_{\alpha,\alpha'}^J$
can be found in Ref.~\cite{[Tam65]}, Eq.~(40) and
Ref.~\cite{[Bug89]}, Eq.~(32).
Here, the coupling potential is obtained by decomposing
the WS  potential
into spherical multipoles up to 12.

We
ensure that enough  daughter
states  are considered for proper convergence.
In practice, we must include some energetically forbidden states.
These states  do not directly
contribute to the width, but do affect the solution.
Furthermore, we  assume that the daughter
nucleus is left in its ground-state rotational band and the
deformation is unchanged during the decay process.
To normalize the cluster radial functions, we use
a method  \cite{[Gya71]} which became known as
``exterior complex scaling'' \cite{[Si79]}.

The description of very narrow proton resonances is a
challenging task due to dramatically different energy scales
of $E_0$ and $\Gamma$. Indeed, while the energies of
single-proton resonances are of the order of 1\,MeV,
their widths can be as small as 10$^{-22}$\,MeV.
This calls for unprecedented numerical accuracy.
In this work,
we apply the piece-wise perturbation method \cite{[Ixa84]}
 generalized to the
coupled-channels case. The calculations are performed
 in extended precision arithmetic.
 The details of the numerical procedure employed are given in
Ref.~\cite{[Kru99a]}.
 As a
check on the calculated widths, we also calculate
 the width from the
probability current expression \cite{[Hum61]}
$\Gamma=\sum_\alpha \Gamma_\alpha$,
where the channel width is
\begin{equation}
\label{partcur}
\Gamma_\alpha
^J=i{{\hbar^2}\over{2\mu}}{{{u_\alpha^{J\prime}}^*(r)u_\alpha^J(r)
-u^{J\prime}_\alpha(r)u_\alpha^{J*}(r)}\over{\sum_{\alpha^{\prime}}\int_0^{r}
|u_{\alpha'}^J(r')|^2dr'}}.
\end{equation}
The
agreement between
the two methods
is  always  better than $0.1\%$.
It should be noted that $\Gamma$ is independent of
$r$. For narrow and isolated resonances,
one can {\em approximate}  the exact Eq.\ (\ref{partcur}) by the
R-matrix expression, as was done in
Refs.~\cite{[Mag98],[Mag99]}.

One  limit of Eq.~(\ref{cc}) is the degenerate case
in which  $Q_{J_d}$ =$Q_{p}$ for all values of $J_d$.
This is the {\em adiabatic approximation} discussed in
 Refs.~\cite{[Tam65],[Bar64]}.
  It is easy to check
 that in the adiabatic limit
the set of new wave functions
\begin{equation}\label{adiabatic}
   u_{JK\,j_pl_p} = \sqrt{2} \ (-1)^{K+J}\sum_{J_{d}}
C^{J_d0}_{j_pK,J-K} u^J_{J_{d}l_pj_p},
\end{equation}
with $|K|$$\le$$j_p$
is also a solution of (\ref{cc}) with an eigenvalue $Q_0$.
For $\Omega$=$K$=$J$, the wave function
$\Psi_{\Omega}=\sum_{j_pl_p} \frac{u_{\Omega\Omega j_pl_p}(r)}{r}{\cal
Y}_{l_pj_p\Omega}$
represents the intrinsic single-particle Nilsson wave function
with the angular momentum projection on the symmetry axis $\Omega$.
As seen from Eq.~(\ref{adiabatic}),  the strongly-coupled intrinsic state
contains contributions from
all the cluster wave functions
corresponding to {\em different core states.}
Another property of  the adiabatic limit is
the existence of  solutions with
$J$$\ge$$\Omega$. Since, as discussed by Tamura
\cite{[Tam65]}, there is no dynamic coupling between the
angular momentum of the proton and that of the daughter
nucleus (the daughter nucleus is perfectly inert during
the proton emission), there exist infinitely many solutions
obtained by combining $\bbox{j}_p$ and $\bbox{J}_d$.
Since the core states are degenerate, all the solutions with
$J$$\ge$$\Omega$ are degenerate as well.

Let us discuss the results of our calculations.
Since for the very proton-rich nuclei considered in this work
practically nothing is known about their spectra,
we parameterize the ground-state band
of the daughter nucleus as $E_J = \kappa J(J+1)$ and fix
$\kappa$ to the experimental value of $E_{2^+}$ (or to the
value taken from systematic trends).
In the limit of infinite moment of inertia ($\kappa\rightarrow 0$) one
reaches the adiabatic limit.

\begin{figure}[htb]
  \begin{center}
  \leavevmode
\epsfxsize=8cm
    \epsfbox{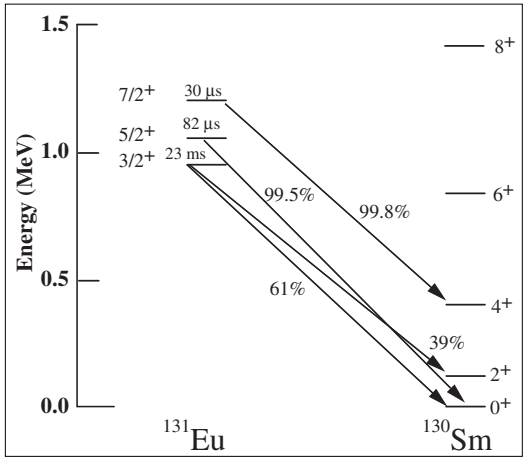}
    \caption{Proton emission from the  [411]3/2
  rotational band in \nuc{131}{Eu}
    calculated in the non-adiabatic model.  Half-lives for proton emission
    and branching ratios are indicated.}
    \label{fig:band}
  \end{center}
\end{figure}

The presence of the  rotational spectrum in the daughter nucleus
gives rise to
 rotational bands in the parent nucleus built upon   the
 $J$=$\Omega$ band-head.
 Figure~\ref{fig:band} shows the calculated rotational band in
\nuc{131}{Eu} built
 upon the  $J=3/2^{+}$ level (associated with the [411]3/2 Nilsson orbital).
  For the energy of the $2^+$ state
  in the daughter nucleus \nuc{130}{Sm} we took the experimental value
\cite{[Son99]}.
 The $J=5/2$ and $7/2$ levels follow very closely the
expected $J(J+1)$ spacing; the small deviations
are due to the Coriolis coupling.

The calculation of branching ratios (b.r.) is straightforward in the
non-adiabatic formalism.  The partial width corresponding to the
$J$$\rightarrow$$J_d$ decay is given by
$\Gamma(J\rightarrow J_d) = \sum_{j_pl_p} \Gamma^J_{J_dj_pl_p}$
It is seen that  the proton-emission lifetimes and b.r.\ 
change with $J$. 
Of course, at the low energies shown in Fig.~\ref{fig:band},  rotational states
decay by emitting gamma radiation ($\Gamma_\gamma$$\gg$$\Gamma_p$).
(The competition with proton emission is expected to take place at higher
energies/spins.)
The 3/2$^+$$\rightarrow$0$^+$ decay  is  given by the very small
$d_{3/2}$
 component. The large branching to the 2$^+$
state is due to the dominating  $d_{5/2}$ partial wave.
(This also explains the fact that the 5/2$^+$ level decays predominantly to the
0$^+$
ground state.)
It is worth noting that
the $s_{1/2}$ component,   not
allowed in the adiabatic approach  due to
 $K$-conservation, also contributes to the
3/2$^+$$\rightarrow$2$^+$
 transition.
Although $K$
is not conserved in the non-adiabatic approach, one  can decompose
the wave function into states (\ref{adiabatic}) with definite
  $K$.
In most cases, we find that one $K$-component dominates the wave function.
Consequently,  the Nilsson labeling convention can still be used.

Transitions to excited daughter states, $K$$\rightarrow$$J_d$,
may also be approximated in the strong coupling
framework  \cite{[Son99],[Bug89]}.  In this case
the angular momentum conservation is guaranteed by the presence
of the geometric factor  $(C_{KK, K-K}^{J_d0})^2$. In addition,
the $Q_p$ value is  adjusted to $Q_{J_d}$.

As shown previously in Ref.~\cite{[Ryk99]}, at large deformations
our calculations show a very
small dependence on $\beta_2$ and $\beta_4$.
This is because the spherical decomposition  of the corresponding Nilsson
orbitals
varies little in this regime, and there are no crossings between the levels
of interest.
The
uncertainty due to the $\beta_{2}$ value is usually much
smaller than the
experimental uncertainty in the proton energy.
Table~I shows predicted half-lives and b.r.\ 
for \nuc{131}{Eu} and  \nuc{141}{Ho}.  For
\nuc{131}{Eu} we take $\beta_{2}=0.32$ and for \nuc{141}{Ho}
($\beta_{2}=0.29,  \beta_{4}=-0.06$) \cite{[Ryk99]}.  Spectroscopic factors
have been
estimated in the independent-quasiparticle picture.
Note that the  $1/(\Omega+1/2)$ coefficient multiplying
the BCS values of $u^2$,  assumed in Ref.~\cite{[Ryk99]},  is no longer
present.
 \begin{table}[btp]
  \begin{tabular}{lllccc}
  &  Orbital & $u^{2}$ & $\tau_{1/2}$ &
    b.r.\ (nad) & b.r.\ (ad) \\ \hline
     &     $[411]\half{3}$ & 0.71  & 34.0 ms  & 39\%  & 37\%  \\
  \nuc{131}{Eu}  & $[413]\half{5}$ & 0.52  & 184  ms  &  7\% & 2\%  \\
  &      $[532]\half{5}$ & 0.48  & 3.90 s   & 52\%  & 38\%  \\
    &                     &      &   {\bf 17.8(19) ms} & {\bf 24(5) \%} &
\\[2mm]
  \nuc{141}{Ho}  &   $[523]\half{7}$ & 0.84  & 19.1 ms      & 6\% & 3\%  \\
                  &                    &      &  {\bf 3.9(5) ms} & & \\
  \nuc{141m}{Ho}       &  $[411]\half{1}$ & 0.70  & 3.3 $\mu$s  & 1\%  & 1\% \\
                  &                    &      &  {\bf 8(3)}
\boldmath{$\mu$}{\bf s}  & &
     \end{tabular}
  \caption{Half-lives and branching ratios (b.r.)  to the $J_d^+$=$2^+$
  state
  for deformed
  proton resonances in  \nuc{131}{Eu} and \nuc{141}{Ho}
  calculated in  the non-adiabatic (nad) and
    adiabatic (ad) formalism.  The experimental  values (shown in boldface)
    are taken from Refs.~\protect\cite{[Son99],[Ryk99]}. The energy
    of the 2$^+$ state in $^{130}$Sm and $^{140}$Dy was assumed to be
    120\,keV and 160\,keV, respectively.}
\end{table}

Considering both the half-life and b.r.,
the ground state of \nuc{131}{Eu} is consistent with the
$[411]3/2$ assignment. This result agrees with the
analysis of Ref.~\cite{[Son99]} and is at variance with the suggestion
by Maglione {\em et al.} who assigned the [413]5/2 orbital as a ground
state of \nuc{131}{Eu}.  The very small b.r.\ for the [413]5/2  orbital
results from the fact that both the 5/2$^+$$\rightarrow$0$^+$
and 5/2$^+$$\rightarrow$2$^+$ transitions go via the
$d_{5/2}$ component which constitutes about 4\% of the wave function.
 On the other hand,  for the
yet-unobserved [532]5/2 state, the 5/2$^-$$\rightarrow$0$^+$ transition
goes via the tiny (0.1\%) $f_{5/2}$  wave,  while the
5/2$^-$$\rightarrow$2$^+$ decay
is dominated by the  $f_{7/2}$ component (17\%)
and the $K$-forbidden $p_{3/2}$ wave, which appears due to the Coriolis
coupling.
This results in
the huge branching predicted for this state.

For \nuc{141}{Ho},
based on calculated  half-lives,  we can assign the
 $[523]7/2$ level to the ground
state  and $[411]1/2$ to the excited  state.
We note that these assignments are
identical to our previous assignment in Ref.~\cite{[Ryk99]}, although we
are now using the non-adiabatic formalism and have changed the optical model
parameters.  These assignments also agree with those proposed in
Refs.~\cite{[Dav98],[Mag99]}.

\begin{figure}[htb]
  \begin{center}
\leavevmode
\epsfxsize=8cm
    \epsfbox{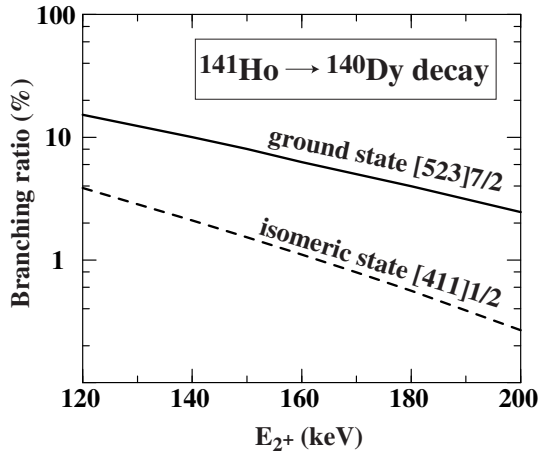}
    \caption{Predicted branching ratio
    for the proton transition from $J^\pi$=7/2$^-$ and  1/2$^+$
    states in \nuc{141}{Ho}
     to  the $2^+$ state in \nuc{140}{Dy} as a
      function of $E_{2^+}$. The expected value of $E_{2^+}$ is around
160\,keV.}
    \label{fig:2+}
  \end{center}
\end{figure}

For \nuc{140}{Dy}, the energy of the $2^+$ state is experimentally
unknown, but it can be estimated
from systematic trends. For instance, according to the $N_pN_n$ scheme,
one obtains the value $E_{2^+}$=160\,keV \cite{[Zam99]}, which was adopted
in the calculations displayed  in Table~I.
Figure~\ref{fig:2+} shows the
expected b.r.\ to the $2^+$ state as a function of
$E_{2^+}$ for both proton-emitting states in  \nuc{141}{Ho}. For the [523]7/2
ground state, the predicted b.r.\ is still of the order of a few percent even
at relatively large values of $E_{2^+}$, and this offers good prospects for
its experimental observation. On the other hand, the b.r.\ for the
isomeric [411]1/2 state is lower by an order of magnitude.
In \nuc{131}{Eu} the experimental uncertainty in the $Q_{p}$ value
leads to an uncertainty of $+30\%$ / $-22\%$ for all three levels.  The
branching ratios vary by less than $3\%$.

In conclusion, we applied a non-adiabatic formalism,  based on the
coupled-channel Schr\"odinger equation with outgoing wave boundary conditions,
to describe very narrow proton resonances in deformed nuclei. The
non-adiabatic model takes into account the fact that the daughter nucleus
has a  finite moment of inertia.
Our calculations are consistent with
the experimental data for  the best deformed proton emitters known so far:
$^{131}$Eu and $^{141}$Ho. As shown in Table~I,
the adiabatic approximation
gives rise to an underestimation of branching ratios
sometimes by a factor  2-3.
According to our predictions, there is a good chance to study  experimentally
the fine structure in the proton decay of $^{141}$Ho.

\acknowledgments
Useful discussions with Krzysztof Rykaczewski are gratefully
acknowledged.
This research was supported in part by
the U.S. Department of Energy
under Contract Nos.\ DE-FG02-96ER40963 (University of Tennessee),
DE-FG05-87ER40361 (Joint Institute for Heavy Ion Research),
and
DE-AC05-96OR22464 with Lockheed Martin Energy Research Corp.\ (Oak
Ridge National Laboratory),
and Hungarian OTKA Grant No. T026244 and No. T029003.


\begin{thebibliography}{10}

\bibitem{[Hof95b]}
{S. Hofmann, Radiochim. Acta {\bf 70/71}, 93 (1995)}.

\bibitem{[Woo97]}
{P.J. Woods and C.N. Davids, Ann. Rev. Nucl. Part. Sci. {\bf 47}, 541 (1997)}.

\bibitem{[Dav98]}
{C.N. Davids {\em et al.},  Phys. Rev. Lett. {\bf 80}, 1849 (1998)}.

\bibitem{[Bat98]}
{J.C. Batchelder {\em et al.},  Phys. Rev. {\bf C57}, R1042
  (1998)}.

\bibitem{[Ryk99]}
{K. Rykaczewski {\em et al.},
  Phys. Rev. C {\bf 60}, 011301 (1999)}.

\bibitem{[Bin99]}
{C.R. Bingham {\em et al.},
  Phys. Rev. C {\bf 59}, R2984 (1999)}.

\bibitem{[Son99]}
{A.A. Sonzogni {\em et al.},  Phys. Rev. Lett. {\bf 83},
  1116 (1999)}.

\bibitem{[Abe97]}
{S. \AA berg, P.B. Semmes, and W. Nazarewicz, Phys. Rev. C {\bf 56}, 1762
  (1997)}.

\bibitem{[Bug89]}
{V.P. Bugrov and S.G. Kadmenski\u{\i}, Sov. J. Nucl. Phys. {\bf 49}, 967
  (1989);
S.G. Kadmenski\u{\i} and V.P. Bugrov, Phys. Atomic Nuclei {\bf 59}, 399
  (1996)}.

\bibitem{[Kad73]}
{S.G. Kadmenski\u{\i}, V.E. Kalechtis, and A.A. Martynov, Sov. J. Nucl. Phys.
  {\bf 14}, 193 (1972); S.G. Kadmenski\u{\i} and V.G. Khlebostroev,
Sov. J. Nucl. Phys. {\bf 18}, 505 (1974)}.

\bibitem{[Fer97]}
{L.S. Ferreira, E. Maglione, and R.J. Liotta, Phys. Rev. Lett. {\bf 78}, 1640
  (1997)}.

\bibitem{[Mag98]}
{E. Maglione, L.S. Ferreira, and R.J. Liotta, Phys. Rev. Lett. {\bf 81}, 538
  (1998)}.

\bibitem{[Mag99]}
{E. Maglione, L.S. Ferreira, and R.J. Liotta, Phys. Rev. C {\bf 59}, R589
  (1999)}.

\bibitem{[Tal98]}
{P. Talou, N. Carjan, and D. Strottman, Phys. Rev. {\bf C58}, 3280 (1998)}.

\bibitem{[Hum61]}
{J. Humblet and L. Rosenfeld, Nucl. Phys. {\bf 26}, 529 (1961)}.


\bibitem{[Tam65]}
{T. Tamura, Rev. Mod. Phys. {\bf 67}, 679 (1965)}.

\bibitem{[Cwi87]}
{S. \'Cwiok, J. Dudek, W. Nazarewicz, J. Skalski, and T. Werner, Comput. Phys.
  Commun. {\bf 46}, 379 (1987)}.

\bibitem{[Che67a]}
{V.A. Chepurnov, Yad. Fiz. {\bf 6}, 955 (1967); Sov. J. Nucl. Phys. {\bf 7},
  715 (1968)}.

\bibitem{[Bec69]} {F.D. Becchetti, Jr. and G.W. Greenlees,
Phys. Rev. {\bf 182}, 1190 (1969)}.

\bibitem{[Naz90]}
{W. Nazarewicz, M.A. Riley, and J.D. Garrett, Nucl. Phys. {\bf A512}, 61
  (1990)}.

\bibitem{[Dud81]}
{J. Dudek, Z. Szyma\'nski, and T. Werner, Phys. Rev. {\bf C23}, 920 (1981)}.

\bibitem{[Nil55]}
{S.G. Nilsson, Mat. Fys. Medd. Dan. Vid. Selsk. {\bf 29}, No. 16 (1955)}.

\bibitem{[Gya71]}
{B. Gyarmati and T. Vertse, Nucl. Phys. {\bf A160}, 523 (1971)}.

\bibitem{[Si79]} {B. Simon, Phys. Lett. {\bf 73A}, 211 (1979)}.

\bibitem{[Ixa84]}
{L.Gr. Ixaru, {\em Numerical Methods for Differential Equations}, (Reidel,
  Dordrecht-Boston-Lancaster, 1984)}.

\bibitem{[Kru99a]}
{T. Vertse, A.T. Kruppa, L.Gr. Ixaru, and M. Rizea, in preparation}.

\bibitem{[Bar64]}
{B.C. Barrett, Nucl. Phys. {\bf 51}, 27 (1964)}.

\bibitem{[Zam99]}
{N.V. Zamfir, private communication, 1999}.

\end{thebibliography}

\end{document}